\documentclass[12pt]{article}

\usepackage[latin2]{inputenc}
\usepackage{amsmath,amsfonts,amssymb,amsthm}
\usepackage{graphicx}
\usepackage{indentfirst}
\usepackage[usenames]{xcolor}
\usepackage{enumerate}
\usepackage[overload]{empheq}
\usepackage{hyperref}

\usepackage[page,title]{appendix}

\usepackage{geometry}
\def\ol{\overline}

\def\u{\mathrm{u}}
\def\uLS{\mathrm{u_{LS}}}
\def\uHS{\mathrm{u_{HS}}}
\def\uA{\mathrm{u_A}}
\def\uNA{\mathrm{u_{NA}}}

\def\iLS{i_{\mathrm{LS}}}
\def\iHS{i_{\mathrm{HS}}}
\def\iA{i_{\mathrm{A}}}
\def\iNA{i_{\mathrm{NA}}}
\def\IHS{\mathrm{I_{HS}}}
\def\ILS{\mathrm{I_{LS}}}
\def\IA{\mathrm{I_{A}}}
\def\INA{\mathrm{I_{NA}}}
\def\IE{\mathrm{I_E}}

\def\cHS{\mathrm{c_{HS}}}
\def\cA{\mathrm{c}_A}

\def\RDLS{\mathrm{RD}_{LS}}
\def\RDA{\mathrm{RD}_A}
\def\RDNA{\mathrm{RD}_{NA}}
\def\J{\mathbf J}
\def\SWN{\mathrm{SW}_{N}}
\def\SWM{\mathrm{SW}_{M}}

\newtheorem{claim}{Claim}

\newtheorem{lemma}{Lemma}

\newtheorem{corollary}{Corollary}

\theoremstyle{definition}

\newtheorem{example}{Example}
\newtheorem{remark}{Remark}

\title{The interplay between migrants and natives\\ as a determinant of migrants' assimilation: \\
A coevolutionary approach}

\author{Jakub Bielawski\footnote{Corresponding author} \footnote{Cracow University of Economics, ul. Rakowicka 27, 31-510 Krak\'ow, Poland, email: bielawsj@uek.krakow.pl} \and Marcin Jakubek\footnote{Institute of Economics, Polish Academy of Sciences, ul. Nowy \'Swiat 72, 00-330 Warszawa, Poland, email: mjakubek@inepan.waw.pl}}

\begin{document}

\maketitle

\begin{abstract}
We study the migrants' assimilation, which we conceptualize as forming human capital productive on the labor market of a developed host country, and we link the observed frequent lack of assimilation with the relative deprivation that the migrants start to feel when they move in social space towards the natives. In turn, we presume that the native population is heterogenous and consists of high-skill and low-skill workers. The presence of assimilated migrants might shape the comparison group of the natives, influencing the relative deprivation of the low-skill workers and, in consequence, the choice to form human capital and become highly skilled. To analyse this interrelation between assimilation choices of migrants and skill formation of natives, we construct a coevolutionary model of the open-to-migration economy. Showing that the economy might end up in a non-assimilation equilibrium, we discuss welfare consequences of an assimilation policy funded from tax levied on the native population. We identify conditions under which such costly policy can bring the migrants to assimilation and at the same time increase the welfare of the natives, even though the incomes of the former take a beating.
\end{abstract}

\noindent{\it Keywords}: Coevolutionary economics, Migrants' assimilation, Well-being of native inhabitants, Relative deprivation

\noindent{\it JEL classification}: C73, D63, F22, J31, J61

\section{Introduction}

We define the migrants' assimilation as forming the human capital (e.g. learning a language) usable on the labor market of a developed host country, which increases their productivity and, in turn, earnings.\footnote{In such a constrained definition, we follow \cite{Borjasetal1992}, who study the assimilation in context of forming ``location-specific human capital''.} Still, much evidence is found that the migrants do not assimilate much, thus their incomes remain low (see, for example \cite{ChiswickMiller2005, Cutler2008, McManus1983, ShieldsPrice2002}). The economic literature gives some possible reasons for this - seemingly irrational - lack of assimilation.\footnote{For example, \cite{Lazear1999} finds the low assimilation to be a consequence of migrants living in concentrated communities. \cite{BeizinMoizeau2017} study the role of culture preservation in urban segregation and lack of socioeconomic integration.} We chose to follow here a strand of literature \cite{FanStark2007, StarkJakubek2013, StarkJakubekSzczygielski2018} that links the low assimilation with the relative deprivation that the migrants feel in comparison with the (richer) natives.

Namely, we start with the presumption that income comparisons matter to the individuals (be it migrants or natives), and that these are mostly upward comparisons which lower the well-being. To quantify the effect of this comparisons in the individuals' preferences, we use the index of relative deprivation.\footnote{The measure was proposed by \cite{Yitzhaki1979} and further axiomatized by \cite{EbertMoyes2000} and \cite{BossertDAmbrosio2006}. Vast empirical evidence supports the significance of relative deprivation in people's well-being; see, for example, \cite{Clark2008, Luttmer2005, WalkerSmith2002}.}

Next, following \cite{Akerlof1997} theory of social proximity and group affiliation, we draw a link between assimilation of migrants and their move in social space toward the natives. This move increases the importance of natives as a comparison group for a migrant, which intensifies the strength of income comparisons between her and the (richer) natives. This intensification in relative deprivation might decrease the benefits from assimilation, even if the absolute income of the migrant rises in the process.

Still, the papers of \cite{FanStark2007, StarkJakubek2013, StarkJakubekSzczygielski2018} treated the behavior and incomes of the counterparts of the migrants' comparisons - the natives - as constant and given exogenously. Here we try to correct for this lacuna by proposing a behavior model that takes into account both sides of the aforementioned comparisons between migrants and natives. In the chosen approach we use a system of replicator dynamics to define a coevolutionary game which describes the interrelated behavior of migrants and natives with respect to choices concerning the assimilation (for the migrants) and human capital formation (for the natives).\footnote{Although the coevolutionary approach is sometimes used to model choices on the border of economics and biology (see, for example, \cite{Noailly2008}), its application to the subject of international migration is scarce. A model of coevolution of natives and migrants can be found in \cite{BarreiraDaSilvaRocha2013}, where a system of two replicator dynamics equations was used to describe the formation of nationalistic attitudes among natives and assimilation of immigrants. However, the model presented in \cite{BarreiraDaSilvaRocha2013} is not fully coevolutionary, as individuals do not derive payoff from meetings with members of their own population. Moreover, the approach taken in \cite{BarreiraDaSilvaRocha2013} does not include the relative deprivation effect.}

Specifically, we presume that the factors that affect the migrants' well-being (payoffs in the game) are their earnings, relative deprivation, and cost of assimilation. The migrants face a choice between assimilating to the mainstream culture, in which case their productivity and earnings increase, but they bear then the costs of assimilation and of intensified comparisons with the natives.

In turn, for the natives the matter of choice is the formation of human capital and consequently to become either a high-skilled worker or remain low-skilled. The natives' well-being (payoffs in the game) depend on their earnings, relative deprivation and cost of human capital formation. To include the positive externality to the productivity of the economy brought by the presence of high-skilled workers, we assume that the earnings of natives and assimilating migrants depend positively on the fraction of high-skill workers in the native population.\footnote{For evidence on a positive effect of human capital spillovers on overall labor productivity and / or wages see, for example, \cite{Rauch1993} and \cite{Moretti2004}. Still, for simplicity, we include in the model an economy-wide spillover effect, rather than local effects that are found by the aforementioned studies.} 

Although in this paper we use a relative deprivation index that assumes upward comparisons, and we assume that the earnings of migrants, assimilating or not, remain lower than those of low-skilled natives, we identify a relative deprivation effect of assimilation on the well-being of natives that is close to the idea of \cite{StarkBielawskiJakubek2014}. Namely, because the measure of relative deprivation depends on the size of the group that an individual compares her income with, entry of the assimilating migrants into the social space of a native may influence her relative deprivation and, in turn, her well-being. Therefore, we consider a possibility that the host-country government might be interested in shaping the assimilation process by means of an assimilation policy. The policy, funded from a tax on the earnings of natives, high- and low-skilled alike, is assumed to work to decrease the cost of assimilation. The main aim of the paper is to analyze how the assimilation policy of the host country affects the equilibria of the evolutionary game.

Before proceeding, a comment is in order. In the empirical literature there is an ongoing discussion of the influence of low-skill migration on the wages and / or employment of low-skilled natives, with findings ranging from highly negative (see, for example, \cite{Borjas2017}) to neutral or even positive (\cite{FogedPeri2016}). To keep the coevolutionary model simple, we do not include directly the effect of appearance of migrants on the earnings of the low-skill natives. However, when discussing the assimilation policy, we show that it is possible that the well-being of the natives increases even if their earnings are diminished as a result of collecting tax to fund the policy.

The main results of the analysis are as follows. First, we find that the decisions on assimilation of migrants and the human capital formation of the natives are interrelated. Second, the group of migrants can become stuck in the non-effective equilibrium with no assimilation, if unaffected by government policy. Third, we identify conditions under which an assimilation policy, funded from tax levied on the natives' earnings, can bring the group of migrants to full assimilation and, moreover, increase the well-being of migrants and natives alike, compared to the no-assimilation outcome.

As a starting point of the analysis, in the next Section we present a simple dynamical model of natives' behavior in a closed-to-migration economy that serves as a benchmark case. In Section 3 we add the migrants into the picture, and we construct a coevolutionary system of equations describing the assimilation behavior of migrants and skill formation of natives, and we introduce an assimilation policy. We then conduct stability analysis of the dynamical system, and we discuss how the assimilation policy affects the equilibria of the evolutionary game. In Section 4 we analyze the welfare effects of the policy-enhanced assimilation process. Section 5 concludes.


\section{Dynamics of the natives}

As a benchmark model, we consider in this Section a closed-to-migration economy. Let there be a country with continuous population of size $N$ of native inhabitants. The workforce of the country is heterogenous: a worker can be low-skilled or high-skilled. The fraction of high-skill natives is denoted by $q$ (the fraction of low-skill natives is then $1-q$). Low-skilled workers enjoy a lower level of earnings than high-skilled workers, but they do not bear costs associated with education. High-skilled workers need to cover a cost of forming their human capital, but they are earning higher wage and also they bring a positive externality to the productivity of the economy. To measure this externality, we assume that the incomes of both high- and low-skilled workers are composed from a "base salary" and an added factor depending on the fraction of high-skill natives. Namely, a high-skilled worker's income, $\iHS(q)$, is equal to $\IHS$ as a base salary plus $q \cdot \IE$, while a low-skilled worker's income, $\iLS(q)$, is equal to $\ILS$ as a base salary plus $q \cdot \IE$, where $\IHS > \ILS > 0$ and $\IE>0$ is the parameter measuring the strength of the externality. We assume that the cost of forming human capital to become a high-skill worker also depends on the fraction of high-skill workers, and amounts to $q \cdot c_{HS}$, where $c_{HS}>0$.\footnote{In other words, the larger the share of high-skill workers in the population, the more effort is needed to be perceived as one.}

To introduce social preferences into the model, we assume that the individuals experience relative deprivation, namely an individual senses dissatisfaction if other individuals earn more than her.
%
The relative deprivation of an individual is defined by means of the index of relative deprivation, namely as a fraction of those whose incomes are higher
than her income times their mean excess income.  In our case, as low-skill workers have lower incomes than the high-skilled ones, the former are relatively deprived. The relative deprivation of a low-skill native is
\begin{align*}
 \RDLS(q) &:= q \cdot \left[(\iHS(q) - \iLS(q) \right] \\
 &= q \cdot \left[(\IHS + q \cdot \IE) - (\ILS + q \cdot \IE) \right] \\
 &= q \cdot (\IHS - \ILS).
\end{align*}

Every native individual derives utility from her income. Moreover, high-skill natives bear the cost of forming human capital, while low-skill natives are concerned about relative deprivation. Thus the utility of a high-skill native is
\begin{align}
\begin{split}\label{uHS1}
 \uHS(q) &:= (1-\beta)\cdot \iHS(q) - q \cdot \cHS \\
 &=(1-\beta) \cdot (\IHS + q \cdot \IE) - q \cdot \cHS,
\end{split}
\end{align}
and the utility of a low-skill native has the form
\begin{align}
\begin{split}\label{uLS1}
 \uLS(q) &:= (1-\beta) \cdot\iLS(q) - \beta\cdot  \RDLS(q) \\
 &=(1-\beta) \cdot (\ILS + q \cdot \IE) - \beta \cdot q \cdot (\IHS - \ILS),
\end{split}
\end{align}
where $\beta \in (0,1)$ describes the intensity of the concern of an individual about being relatively
deprived, with the complementary weight $1-\beta$ defining the utility brought from the level of absolute income.

We describe the evolution of the proportion of high-skill natives, $q$, using the replicator dynamics equation:

\begin{equation}\label{evolution:N}
 \dot{q} = q \cdot [\uHS(q) - \u_N(q)] = q \cdot (1-q) \cdot [\uHS(q) - \uLS(q)],
\end{equation}
where $\u_N(q)$ denotes average utility of natives, that is $\u_N(q) = q \cdot \uHS(q) + (1-q) \cdot \uLS(q)$.
The replicator dynamics reflects the fact that the fraction of high-skill natives, $q$, increases as long as the utility of a high-skill native is higher than the average utility of the
population.

Solving equation \eqref{evolution:N} for steady states is equivalent to $q=0$ or $q=1$ or $\uHS(q)=\uLS(q)$. The last equation has only one solution given by
\begin{equation}\label{q*}
 q^* = \frac{(1-\beta) \cdot (\IHS-\ILS)}{\cHS-\beta \cdot (\IHS-\ILS)}.
\end{equation}
The solution $q^*$ is internal (i.e. $q^* \in (0,1)$) if and only if
\begin{equation}\label{stationary:N}
 \cHS > \IHS-\ILS.
\end{equation}

If $\cHS \leqslant \IHS - \ILS$ then the only steady states of the dynamics \eqref{evolution:N} are $q=0$ and $q=1$. In other words in the long run the population of natives consists entirely of low-skill workers ($q=0$) or of high-skill workers ($q=1$). To assure that the model of economy is non-trivial, namely that there exist non-zero fractions of both low- and high-skill natives, for the analysis that follows we assume that the condition \eqref{stationary:N} holds.

Stability of the steady states of the equation \eqref{evolution:N} is summarized below.
\begin{corollary}\label{q*N}
For the dynamics described by equation \eqref{evolution:N}, we have that:
\begin{itemize}
 \item $q=0$ is unstable,
 \item $q=1$ is unstable,
 \item $q=q^*$ is asymptotically stable.
\end{itemize}
\end{corollary}
The proof of Corollary \ref{q*N} is in Appendix \ref{ApxN}.

Thus only $q=q^*$ forms a stable equilibrium in closed-to-migration economy. In other words, starting from any level of fraction of high-skill natives such that $q_0\in (0,1)$, the dynamics tends to $q^*$ as time approaches infinity. 

\begin{example}
The dynamics of natives for the following values of the parameters:
$$
 \begin{array}{lllll}
 \IHS = 1.0, & \hspace*{1cm} & \IE = 0.35 & \hspace*{1cm} & \beta = 0.5, \\
 \ILS = 0.6, & \hspace*{1cm} & \cHS = 0.7,
 \end{array}
$$
is shown on the graph below. In this case only $q^* = 0.4$ is asymptotically stable steady state.
\begin{figure}[ht]
 \centering
  \includegraphics[width=0.8\textwidth]{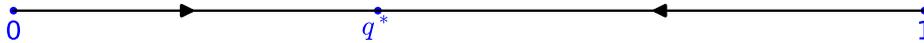}
  \caption{Phase portrait of dynamics \eqref{evolution:N}.}
\end{figure}
\end{example}


\section{Co-evolution of natives and migrants}

In this Section, we introduce a population of migrants to the country of natives. The population of migrants is continuous and of size $M<N$. Every migrant has two available strategies: she may decide to assimilate, that is, learn the language of natives, obtain tools and skills which increase her productivity at the labor market of the host country, or she may remain non-assimilating. We denote by $p$ the fraction of assimilating migrants, with $1-p$ being the fraction of non-assimilating migrants.

The assimilation defined in such a manner is costly; forming the human capital by a migrant is connected with cost $\cA\geq0$. As we are interested in an institutional response of the host country to the appearance of the migrants, we introduce a possibility that the host country implements an assimilation policy, which has a form of an allowance $A$ aimed at reducing the cost of assimilation, such that $\cA > A \geq 0$, where $A=0$ represents the case without an assimilation policy. At the same time the natives bear the cost of this operation, i.e. the income of every native is reduced by $\frac{p \cdot M \cdot A}{N} = p \cdot m \cdot A$, where $m := \frac{M}{N} < 1$. The income of a high-skill native is now $\iHS(p,q) = \IHS + q \cdot \IE - p \cdot m \cdot A$, and the income of a low-skill native $\iLS(p,q)$ equals now to $\ILS + q \cdot \IE - p \cdot m \cdot A$.

An assimilating migrant's income, $\iA(p,q)$, is equal to $\IA$ as a base salary plus $q \cdot \IE$, namely she also benefits from the externality provided by the high-skilled natives. A non-assimilating migrant earns a wage $\iNA(p,q) = \INA$ (no externality from high-skilled natives occurs in her case). We assume that the base salary of assimilating migrants is higher then the income of non-assimilating migrants and at the same time lower then the base salary of low-skill natives after taxation for every possible levels of assimilation of migrants, $p$, and of assimilation allowance $A<\cA$, i.e. $\ILS - m \cdot \cA > \IA > \INA > 0$.\footnote{The assumption on migrants' incomes being lower than those of low-skill natives is, of course, a simplification, however the Eurostat data show that indeed the migrants' median income was even 50\% lower than that of nationals in some EU countries, and that almost half of migrant population in EU-28 was at risk of poverty or social exclusion in 2016 (\cite{Eurostat2018}).}

\subsection{The co-evolutionary system}

To analyse the influence of migrants on the group of natives, we presume that, apart from imposing a possible cost of the assimilation policy, the assimilation of migrants widens the reference group of natives, namely the natives include the assimilating migrants in their reference group. Because the earnings of assimilating migrants remain lower than those of the low-skill workers, the widening of the comparison group decreases the relative deprivation of the latter, which now amounts to
\begin{align*}
 \RDLS(p,q) &:= \frac{q N}{N+p M} \big( \iHS(p,q) - \iLS(p,q) \big)\\
 &= \frac{q}{1+p \cdot m}  \left[(\IHS + q \cdot \IE - p \cdot m \cdot A)
 - (\ILS + q \cdot \IE - p \cdot m \cdot A)\right] \\
 &=\frac{q}{1+p \cdot m} (\IHS - \ILS).
\end{align*}

The utility of low-skilled native is now thus:
\begin{align}
\begin{split}\label{uLS2}
 \uLS(p,q) &= (1-\beta) \cdot \iLS(p,q) - \beta \cdot \RDLS(p,q)\\
 &= (1-\beta) \cdot (\ILS + q \cdot \IE - p \cdot m \cdot A)
 - \frac{\beta \cdot q}{1 + p \cdot m} (\IHS - \ILS),
\end{split}
\end{align}
while, comparing to $\uHS(q)$ defined in Section 2, the utility of a high-skilled native is now affected by the cost of assimilation policy:
\begin{align}
\begin{split}\label{uHS2}
 \uHS(p,q) &= (1-\beta) \cdot \iHS(p,q) - q \cdot  \cHS\\
 &= (1-\beta) \cdot (\IHS + q \cdot \IE - p \cdot m \cdot A) - q \cdot \cHS.
\end{split}
\end{align}

Although the low-skill natives benefit from the externality provided by the high-skilled natives, we assume that the boost in utility caused by this increase in absolute income is smaller than the negative effect of relative deprivation arising from comparisons with the high-skilled workers for every possible level of assimilation of migrants, that is
\begin{equation} \label{IHS-ILSIq}
  \frac{\beta}{1+m} (\IHS - \ILS) > (1-\beta) \cdot \IE.
\end{equation}
From the last inequality and with help of \eqref{stationary:N} we derive the following chain of inequalities:
$$
 \cHS > \frac{\beta}{1+m} \cHS > \frac{\beta}{1+m} (\IHS-\ILS) > (1-\beta) \cdot \IE.
$$
In particular we obtain that
\begin{equation} \label{cHSIq}
 \cHS > (1-\beta) \cdot \IE.
\end{equation}
The inequality \eqref{cHSIq} yields that the positive externality to the productivity of the economy due to the presence of workforce of high-skill natives is not high enough to compensate for the negative effects on utility of natives brought by the cost of becoming high-skilled.

To analyse the utility of migrants in the social space of natives we assume that, apart from the change in earnings of assimilating migrants, the assimilation affects the utilities of migrants in several other dimensions. First, it brings the migrant closer to the natives; in other words, it is impossible to assimilate in economic dimension and at the same time remain disconnected from the society of the natives. In consequence, the reference group of an assimilating migrant consists of the entire population, thus an assimilating migrant experiences relative deprivation from comparing her income with those of the (richer) natives. The relative deprivation of an assimilating migrant equals thus to
\begin{align*}
\RDA(p,q) &:= \frac{N}{N+M} \left[ q \cdot \big( \iHS(p,q) - \iA(p,q) \big) + (1-q) \cdot \big( \iLS(p,q) - \iA(p,q) \big) \right]\\
&=  \frac{1}{1+m} \Big[ q \cdot \big( (\IHS+q\cdot\IE-p\cdot m\cdot A) - (\IA+q\cdot\IE) \big) \\
&\hspace*{1.8cm} + (1-q) \cdot \big( (\ILS+q\cdot\IE-p\cdot m\cdot A) - (\IA+q\cdot\IE) \big) \Big] \\
&= \frac{1}{1+m} \cdot \big[ q \cdot (\IHS - \ILS) + (\ILS - \IA - p \cdot m \cdot A) \big].
\end{align*}

A non-assimilating migrant experiences relative deprivation only from comparing with the assimilating migrants; the natives are not in the reference group of non-assimilating migrant. The relative deprivation of a non-assimilating migrant amounts to
\begin{align*}
\RDNA(p,q) &:= p \cdot \big( \iA(p,q) - \iNA(p,q) \big)\\
 &= p \cdot \big( \IA + q\cdot\IE - \INA \big).
\end{align*}

The utilities of assimilating and non-assimilating migrants amount to, respectively:
\begin{align*}
 \uA(p,q) &= (1-\beta) \cdot \iA(p,q) - (\cA - A) - \beta \cdot \RDA(p,q)\\
 &= (1-\beta) \cdot (\IA + q \cdot \IE) - (\cA - A) \\
 &- \frac{\beta}{1+m} \cdot\big[ q \cdot (\IHS - \ILS) + (\ILS - \IA - p \cdot m \cdot A) \big]
\end{align*}
and
\begin{align*}
 \uNA(p,q) &= (1-\beta) \cdot \iNA(p,q) - \beta \cdot \RDNA(p,q)\\
 &= (1-\beta) \cdot \INA - \beta \cdot p \cdot (\IA + q \cdot \IE - \INA).
\end{align*}

Empirical studies reveal that assimilation increases migrants' incomes \cite{ McManus1983, ShieldsPrice2002, Tainer1988}. Therefore it is reasonable to assume that if there was no assimilation cost then all migrants would assimilate. To this end it is enough to ensure that $\uA(p,q)_{\big| \cA=0} > \uNA(p,q)_{\big| \cA=0}$ (note that if $\cA=0$ then also $A=0$). Observe that the function $\uA(\cdot)$ is increasing with respect to variable $p$, and that the function $\uNA(\cdot)$ is decreasing with respect to variable $p$. Thus for the last inequality to hold for every $p \in [0,1]$ it is sufficient to assume that this inequality holds for $p=0$. Moreover, by \eqref{IHS-ILSIq} we have that the function $\uA(\cdot)$ is decreasing with respect to variable $q$ (and variable $q$ is not relevant in $\uNA(\cdot)$ if $p=0$), thus for the last inequality to hold for every $q \in [0,1]$ it is sufficient to assume that this inequality holds for $q=1$. Therefore, it is sufficient that $\uA(0,1)_{\big| \cA=0} > \uNA(0,1)_{\big| \cA=0}$ holds, which is equivalent to the following inequality:
\begin{equation}\label{no_assimilation_cost}
 (1-\beta) \cdot (\IA + \IE - \INA) > \frac{\beta}{1+m} (\IHS - \IA).
\end{equation}

We describe the co-evolution of proportions of high-skill native in native population, $q$, and of assimilating migrants in migrant population, $p$, using a system of replicator dynamic equations:
\begin{align}[left=\empheqlbrace]
\begin{split}\label{evolution:MN}
 \dot{p} &= p \cdot (1-p) \cdot (\uA(p,q) - \uNA(p,q)),\\
 \dot{q} &= q \cdot (1-q) \cdot (\uHS(p,q) - \uLS(p,q)).
 \end{split}
\end{align}

The system \eqref{evolution:MN} represents the presumption that as long as assimilation brings higher utility that non-assimilation, the proportion of assimilating migrants ($p$) will increase. Likewise, a the proportion of natives investing in human capital ($q$) will grow if the utility of a high-skill native is higher that that of low-skill native.

\subsection{Stability analysis of the system}

The steady states of the system \eqref{evolution:MN} are the solutions of the equations $\dot{p}=0$ and $\dot{q}=0$. Solving this system of equations can be divided into following 9 cases:
\begin{enumerate}[(A)]
\begin{minipage}{0.25\textwidth}
\item $\left\{\begin{array}{l}
p=0\\
q=0
\end{array}\right.
$

\item $\left\{\begin{array}{l}
p=0\\
q=1
\end{array}\right.
$

\item $\left\{\begin{array}{l}
p=1\\
q=0
\end{array}\right.
$
\end{minipage}
\begin{minipage}{0.35\textwidth}
\item $\left\{\begin{array}{l}
p=1\\
q=1
\end{array}\right.
$

\item $\left\{\begin{array}{l}
\uA(p,q) = \uNA(p,q) \\
q=0
\end{array}\right.
$

\item $\left\{\begin{array}{l}
\uA(p,q) = \uNA(p,q) \\
q=1
\end{array}\right.
$
\end{minipage}
\begin{minipage}{0.35\textwidth}
\item $\left\{\begin{array}{l}
p=0\\
\uHS(p,q) = \uLS(p,q)
\end{array}\right.
$

\item $\left\{\begin{array}{l}
p=1\\
\uHS(p,q) = \uLS(p,q)
\end{array}\right.
$

\item $\left\{\begin{array}{l}
\uA(p,q) = \uNA(p,q) \\
\uHS(p,q) = \uLS(p,q)
\end{array}\right.
$
\end{minipage}
\end{enumerate}

Cases (A)--(D) yield the corners of the square $[0,1] \times [0,1]$, i.e. the states $(0,0)$, $(0,1)$, $(1,0)$ and $
(1,1)$, respectively.

The existence of solutions of the equations (E) and (F) within the square $[0,1] \times [0,1]$ is dependent on the value of the cost of assimilation:
\begin{itemize}
\item The system (E) has exactly one solution $(p^*,0)$, where
$$
 p^* = \frac{\frac{\beta}{1+m}(\ILS - \IA) + (\cA - A) - (1-\beta) \cdot (\IA - \INA) }{\beta \left( \frac{m}{1+m} A + \IA - \INA \right)},
$$
such that $p^* \in (0,1)$ if and only if
\begin{align*}[left=\empheqlbrace]
 \cA &> A + (1-\beta) \cdot (\IA - \INA) - \frac{\beta}{1+m} (\ILS-\IA), \\
 \cA &< \left( 1+\frac{m}{1+m}\beta \right) \cdot A + (\IA - \INA) - \frac{\beta}{1+m} (\ILS-\IA).
\end{align*}

\item The system (F) has exactly one solution $(p^{**},1)$, where
$$
 p^{**} = \frac{\frac{\beta}{1+m}(\IHS - \IA) + (\cA - A) - (1-\beta) \cdot (\IA + \IE - \INA) }{\beta \left( \frac{m}{1+m} A + \IA + \IE - \INA \right)},
$$
such that $p^{**} \in (0,1)$ if and only if
\begin{align*}[left=\empheqlbrace]
 \cA &> A + (1-\beta) \cdot (\IA + \IE - \INA) - \frac{\beta}{1+m} (\IHS-\IA), \\
 \cA &< \left( 1+\frac{m}{1+m}\beta \right) \cdot A + (\IA + \IE - \INA) - \frac{\beta}{1+m} (\IHS-\IA).
\end{align*}
\end{itemize}

The existence of solutions of the equations (G) and (H) within the square $[0,1] \times [0,1]$ is ensured by the inequality \eqref{stationary:N}. Case (G) has a solution $(0,q^*)$, where $q^*$ is given by \eqref{q*}, namely, it is the equilibrium of the closed-to-migration economy, and case (H) has a solution $(1,q^{**})$, where
\begin{equation}\label{q**}
 q^{**} = \frac{(1-\beta) \cdot (\IHS-\ILS)}{\cHS- \frac{\beta}{1+m} (\IHS-\ILS)}.
\end{equation}
Notice that
\begin{equation}\label{q*q**}
 q^* - q^{**} = q^* \cdot q^{**} \cdot \frac{\beta}{1-\beta} \cdot \frac{m}{1+m} > 0.
\end{equation}
The decrease of the fraction of high-skill workers in the case of the state $(1,q^{**})$ in comparison with the state $(0,q^*)$ is a consequence of the fact that increase of the fraction of assimilating migrants decreases the relative deprivation of low-skill natives, thus increases the utility of this group. As a result, the gain from becoming high-skill native diminishes, and the equilibrium $q$ shifts down.

The system of equations (I) has two solutions which can be obtained using simple algebra, however the exact formulas for the solutions are somewhat complicated and do not provide any useful information, therefore we do not show these formulas in full form. Still, we will be able to derive some more information regarding the solutions of the system of equations (I) after we determine how the evolution of the system \eqref{evolution:MN} depends on the assimilation policy.

Compared to the simple dynamics of natives (cf. \eqref{evolution:N}), the local stability of the steady states of the system \eqref{evolution:MN} depends on the interplay between the parameters of the model, most notably the value of $A$. However some information on this subject still can be derived, which is the substance of the following corollary.

\begin{corollary}\label{MNs}
For the dynamical system \eqref{evolution:MN} we have that:
\begin{enumerate}
\item the states $(0,0)$, $(0,1)$, $(1,0)$, $(1,1)$ are all unstable,
\item if $p^* \in (0,1)$, then the state $(p^*,0)$ is unstable,
\item if $p^{**} \in (0,1)$, then the state $(p^{**},1)$ is unstable.
\end{enumerate}
\end{corollary}
The proof of Corollary \ref{MNs} is in Appendix \ref{ApxMNs}.\\

Therefore, as in the case of the close-to-migration economy, the states of the system such that $q=0$ or $q=1$ are unstable. The following claim summarizes the stability of states derived as solutions to the cases (A)--(H).

\begin{claim}\label{AS}
Among the states characterized by the cases (A)--(H), only states $(0,q^*)$ and $(1,q^{**})$ can be asymptotically stable.
Specifically:
\begin{itemize}
\item the state $(0,q^*)$ is asymptotically stable if and only if $A < A^*$, where
\begin{align}
\begin{split}\label{A*}
A^* :=\ &\cA + \frac{\beta}{1+m}(\ILS - \IA) - (1-\beta) \cdot (\IA - \INA) \\
 &+ q^* \cdot \left( \frac{\beta}{1+m}(\IHS - \ILS) - (1-\beta) \cdot \IE \right),
\end{split}
\end{align}

\item the state $(1,q^{**})$ is asymptotically stable if and only if $A > A^{**}$, where
\begin{align}
\begin{split}\label{A**}
A^{**} :=\ &\frac{1}{\left( 1+\frac{m}{1+m}\beta \right)} \left[\cA + \frac{\beta}{1+m}(\ILS - \IA) - (\IA - \INA) \right. \\
 &\left. +\: q^{**} \cdot \left( \frac{\beta}{1+m}(\IHS - \ILS) - \IE \right) \right].
\end{split}
\end{align}
\end{itemize}
\end{claim}

\noindent The proof of Claim \ref{AS} is in Appendix \ref{ApxMNeq}.

\begin{corollary}\label{A**A*}
Let $A^*$ be defined in \eqref{A*} and $A^{**}$ be defined in \eqref{A**}. Then
$$
 A^{**} < A^* < \cA.
$$ 
\end{corollary}

\noindent The proof of Corollary \ref{A**A*} is in Appendix \ref{ApxMNeq}.

Claim 1 and Corollary 3 provide the main results of this Section, but before we proceed with describing them, we finalize the stability discussion by characterizing the solutions to the case that was not discussed as yet, that is (I), in the next lemma and corollary that follows.

\begin{lemma}\label{CaseI}
The system of equations (I) has at most one solution within the square $[0,1] \times [0,1]$.  Moreover, if $A>A^*$, the system of equations (I) has no solution within the square $[0,1] \times [0,1]$.
\end{lemma}

\noindent The proof of Lemma \ref{CaseI} is in Appendix \ref{ApxMNeq}.

\begin{corollary} \label{olp,olq}
If $(\ol{p},\ol{q})$ denotes a solution of the system of equations (I) such that $(\ol{p}, \ol{q}) \in [0,1] \times [0,1]$, then:
\begin{enumerate}
\item $q^{**} \leqslant \ol{q} \leqslant q^*$ if and only if $\ol{p} \in [0,1]$,
\item the state $(\ol{p},\ol{q})$ is unstable.
\end{enumerate}
\end{corollary}

\noindent The proof of Corollary \ref{olp,olq} is in Appendix \ref{ApxMNeq}.

Corollary \ref{olp,olq} states that even if there exists a solution to the system (I) that lies inside the square $[0,1] \times [0,1]$, it is surely not stable.

Summing up the results pertaining the stability of solutions to the cases (A)--(I) above, we get that only the states $(0,q^*)$ and $(1,q^{**})$ are candidates to be asymptotically stable. More specifically, if $A \in (A^{**}, A^*)$ then both states $(0,q^*)$ and $(1,q^{**})$ are asymptotically stable. In this case the basins of attraction of $(0,q^*)$ and of $(1,q^{**})$ divide the square $(0,1) \times (0,1)$ into two disjoint subsets. The dependence of the asymptotic stability of the states $(0,q^*)$ and $(1,q^{**})$ on the assimilation allowance $A$ is represented diagrammatically in the following graph.
\begin{figure}[ht]
 \centering
  \includegraphics[width=0.8\textwidth]{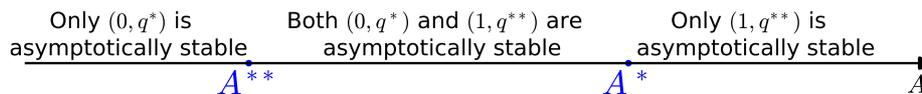}
  \caption{Asymptotic stability of the states $(0,q^*)$ and $(1,q^{**})$ depends on the assimilation allowance.}
\end{figure}

Basing on Claim \ref{AS} and Corollary \ref{A**A*}, we can derive three observations for the economy with migration but without an assimilation policy, which are given in the form of a remark below.
\begin{remark}
~\begin{enumerate}
\item In case $A^{**} > 0$, then if $A=0$, the dynamics is moving all initial states $(p_0, q_0) \in (0,1) \times (0,1)$ towards $(0, q^*)$, i.e. the state where all migrants are non-assimilating.
\item In case $A^{**} < 0 < A^*$, then if $A=0$, some initial states will move towards $(0, q^*)$, i.e. the state where all migrants are non-assimilating, and other initial states will move towards $(1, q^{**})$, i.e. the state where all migrants are assimilating. However, if the fraction of assimilating migrants is initially small enough, then the state $(p_0, q_0)$ will move towards $(0, q^*)$. This type of dynamics is shown in Example \ref{SageExample} below.
\item If $A^* < 0$, then even if $A=0$, the dynamics is moving all initial states $(p_0, q_0) \in (0,1) \times (0,1)$ towards $(1, q^{**})$, i.e. the state where all migrants are assimilating.
\end{enumerate}
\end{remark}

Observe that situation in which $A^* < 0$ requires no assimilation policy for the full-assimilation to be the only asymptotically stable state. However, as such constellation provides no tension in the assimilation process, from now on we focus on the situation in which $A^* > 0$. This last inequality can be expressed by means of a condition on the cost of assimilation: $ A^* > 0 \Leftrightarrow \cA > \ol{\cA}$, where
\begin{align}
\begin{split}\label{olcA}
 \ol{\cA} := &(1-\beta)(\IA - \INA) - \frac{\beta}{1+m}(\ILS - \IA) \\
 &- q^* \cdot \left( \frac{\beta}{1+m}(\IHS - \ILS) - (1-\beta) \cdot \IE \right).
\end{split}
\end{align}
Moreover, note that the factor $\ol{\cA}$ is positive. Indeed, $A^* = \cA - \ol{\cA}$ and by Corollary \ref{A**A*} we have that $A^* < \cA$. These two facts imply that $\ol{\cA} > 0$.

We now provide a numerical solution and a vector field diagram of the system \eqref{evolution:MN} for a set of chosen values of the parameters.

\begin{example} \label{SageExample}
In the graph below we show the dynamics of the the economy with migration for the following values of the parameters:

$$
 \begin{array}{lllll}
 \IHS=1.0, & \hspace*{1cm} & \IE=0.35, & \hspace*{1cm} & m=0.1, \\
 \ILS=0.6, & \hspace*{1cm} & \cHS=0.7, & \hspace*{1cm} & A=0, \\
 \IA=0.53, & \hspace*{1cm} & \cA=0.2, \\
 \INA=0.3, & \hspace*{1cm} & \beta=0.5,
 \end{array}
$$
along with trajectories of some initial states. In this case $A^{**} = -\frac{86}{1425} \approx -0.0604 < 0$ and $A^* = \frac{263}{2200} \approx 0.1195 > 0$. Because $A^{**} < A < A^*$, we have that both states $(0,q^*) = (0,0.4)$ and $(1,q^{**}) = \left( 1,\frac{22}{57} \right)$ are asymptotically stable.

\begin{figure}[ht]
 \centering
  \includegraphics[width=1\textwidth]{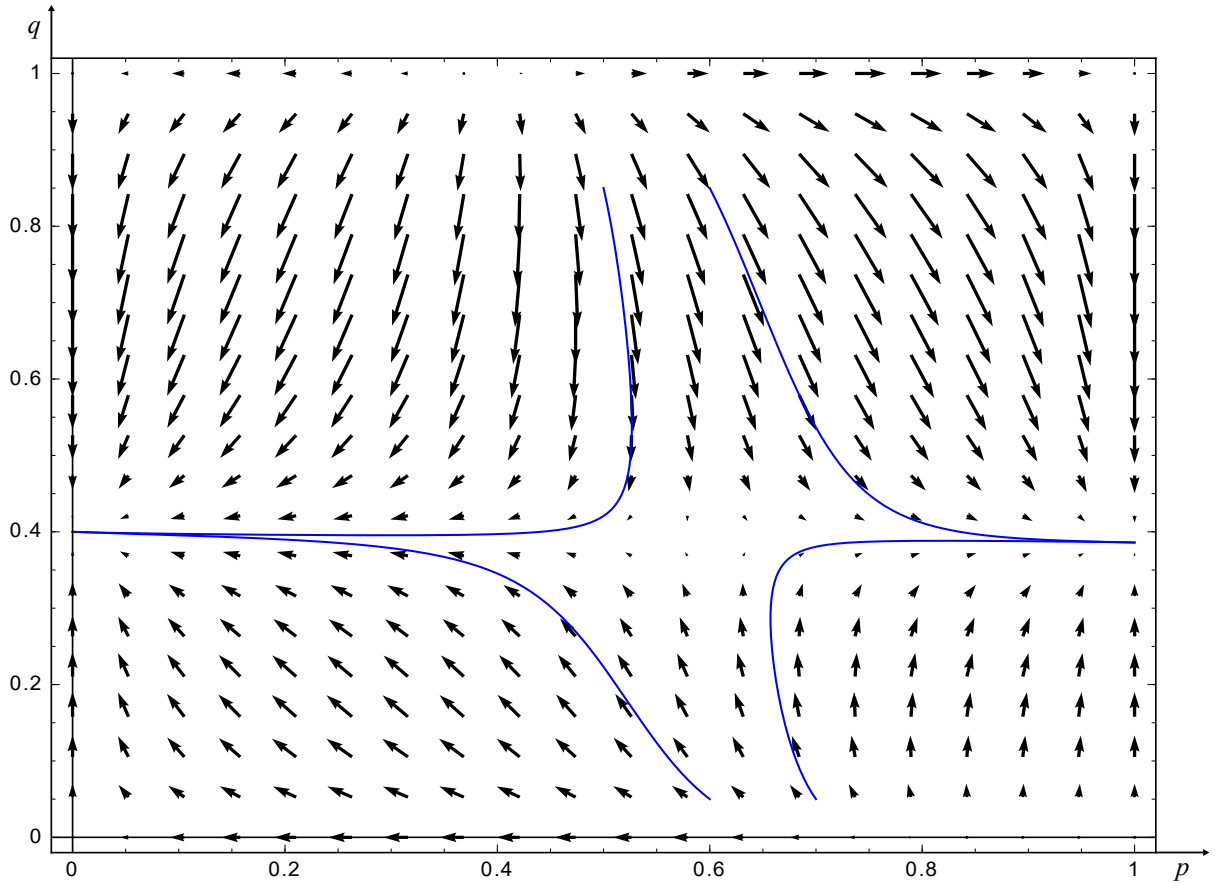}
  \caption{Phase portrait of dynamical system \eqref{evolution:MN}.}
\end{figure}
\end{example}


\section{Social welfare of natives}\label{SW}

In the previous Section we characterized conditions under which it is possible that without an assimilation policy the economy will be at risk of ending up in a configuration, in which all the migrants choose not to assimilate ($p=0$). Assuming that such conditions hold, that is, $A^*>0$ (cf. Remark 1 and discussion that followed), the government might be interested in introducing the policy in order to prevent such an outcome. Thus, in this Section we pose the following question: is it possible that the natives will be better off in an open-to-migration economy in which the assimilation of migrants is triggered by a (costly) assimilation policy in comparison to closed-to-migration economy? We address this problem by providing an analysis of the well-being of the natives measured by an utilitarian social welfare function.

We denote social welfare of natives by $\SWN(p,q,A)$, which we define as the sum of the utilities of natives:
\begin{align}
\begin{split}\label{SWN}
 \SWN(p,q,A) &:= q \cdot N \cdot \uHS(p,q) + (1-q) \cdot N \cdot \uLS(p,q)\\
 &= N \cdot \left[ q \cdot \left( \uHS(p,q) - \uLS(p,q) \right) + \uLS(p,q) \right].
\end{split}
\end{align}

We first note that an equilibrium of the open-to-migration economy $(0,q^*)$ with $A=0$ is equivalent in the native social welfare to an equilibrium $q^*$ in closed-to-migration economy. To see this clearly, it suffices to compare \eqref{uHS1} with \eqref{uHS2} and \eqref{uLS1} with \eqref{uLS2} for $p=0$.

Second, recalling  Claim \ref{AS}, the state $(0,q^*)$ is a stable equilibrium as long as $A < A^*$. Thus if $A \geqslant A^*$ then by Corollary \ref{A**A*} we know that the state $(1,q^{**})$ is the only stable equilibrium in open-to-migration economy. Thus, the minimal level of allowance that enables full assimilation of migrants is $A=A^*$.

Therefore, to answer the question posed at the beginning of this Section, we will compare social welfare of natives in two following equilibria of the economy with migration: $(0,q^*)$ with $A=0$ and $(1,q^{**})$ with $A=A^*$. We now state and prove the following claim.

\begin{claim}\label{full_assimilationN}
The full assimilation of migrants (enhanced by the assimilation policy) is beneficial to natives, i.e. $\SWN(1,q^{**},A^*) > \SWN(0,q^*,0)$, if and only if
\begin{align}\label{SWNcA}
 \cA < \ol{\cA} + q^* \cdot q^{**} \cdot \frac{\beta \cdot \left( \cHS - (1-\beta) \cdot \IE \right)}{(1+m)(1-\beta)^2},
\end{align}
where $\ol{\cA}$ is defined in \eqref{olcA}.
\end{claim}

\noindent The proof of Claim \ref{full_assimilationN} is in Appendix \ref{ApxSW}.

In Claim \ref{full_assimilationN} we identified a condition on the level of the cost of assimilation, for which a successful assimilation policy brings a welfare gain to the natives. If this condition holds, then even if the native population bear the cost of the policy, the overall effect of a decrease in relative deprivation of low-skill natives, brought about by the move of migrants in the social space, is enough to compensate for the loss of utility caused by decrease in absolute income. However, if this condition is not met, then setting in motion the assimilation policy will be harmful to the well-being of the natives. For example, reminding the formula for $q^{**}$ (\ref{q**}), we see that an increase in the fraction of migrants, $m$, tightens the cap on $c_A$ defined on the right-hand-side of (18). Namely, the more migrants are present in the economy, the more stringent the condition that determines the effectiveness of the assimilation policy.

Lastly, we take a look at the migrants' well-being under the policy. We define social welfare of migrants, $\SWM(p,q,A)$, in a similar fashion as in the case of natives:
$$
 \SWM(p,q,A) := p \cdot M \cdot \uA(p,q) + (1-p) \cdot M \cdot \uNA(p,q).
$$

In the following corollary we provide a quite intuitive result, which pertains to the effect of the assimilation policy on the welfare of the migrants.

\begin{corollary}\label{full_assimilationM}
The full assimilation (enhanced by the assimilation policy) is beneficial to migrants (in comparison to no-assimilation).
\end{corollary}

\noindent The proof of Corollary \ref{full_assimilationM} is in Appendix \ref{ApxSW}.

Corollary \ref{full_assimilationM} shows that if there is a need for an assimilation policy to bring the migrants to full assimilation (that is, $A^*>0$), such policy is always beneficial to the migrant population.

In sum, we obtain as the main result of this Section that as long as the condition \eqref{SWNcA} holds, then a costly assimilation policy such that $A=A^*$ is welfare-enhancing for the whole population of the host country, natives and migrants alike.


\section{Conclusions}
We have formulated a simple dynamical model of an open-to-migration developed economy which struggles with the problem of
non-assimilation of migrants. Throughout the paper, we have used a somewhat tailored definition of assimilation, limiting it to the process of acquiring the human capital specific to the labor market of the host country. Still, we admit that the process of assimilation in the economic sphere cannot occur without a parallel move in a social sphere - a move which brings the migrants and natives together and influences the formation of comparison groups of both.

Next, we studied the interplay between the absolute and relative income effects of an assimilation policy, which is targeted at bringing all the migrants to the point of assimilation. We identified conditions under which the policy can increase the welfare of both the natives and migrants, even though it is funded out of the pockets of the former. The conditions for such an outcome to occur hinge on what we have defined for the needs of our model as the migrants' cost of assimilation, which, in turn, can depend on a mixture of many social, cultural and economic factors, characterizing both the migrants as well as the receiving country. In any case, a government pondering an adoption of a costly assimilation-enhancing policy must understand the intricate channels it might affect the economy and social fabric.


\section*{Acknowledgements}

Jakub Bielawski gratefully acknowledges the support of the National Science Centre, Poland, grant no. 2018/02/X/HS4/00673.

\newpage
\begin{appendices}

\section{} \label{ApxN}

\begin{proof}[{\bf Proof of Corollary \ref{q*N}}]

We denote the most right-hand-side of the equation \eqref{evolution:N} by $F(q)$, i.e.
\begin{align*}
 F(q) &= q \cdot (1-q) \cdot [\uHS(q) - \uLS(q)]\\
 &= q \cdot (1-q) \cdot [(1-\beta) \cdot (\IHS - \ILS)
 + \beta \cdot q \cdot (\IHS - \ILS) - q \cdot \cHS].
\end{align*}

The stability of a steady state $r$ of the equation \eqref{evolution:N} depends on the sign of $F'(r)$:
\begin{itemize}
 \item if $F'(r) > 0$ then $r$ is unstable,
 \item if $F'(r) < 0$ then $r$ is asymptotically stable.
\end{itemize}

The derivative of $F$ has the following form
$$
 F'(q) = (1-q) \cdot [\uHS(q) - \uLS(q)] - q \cdot [\uHS(q) - \uLS(q)]
 + q \cdot (1-q) \cdot [\u'_{HS}(q) - \u'_{LS}(q)].
$$
We have that
\begin{enumerate}
 \item $F'(0) = \uHS(0) - \uLS(0) = (1-\beta) \cdot (\IHS - \ILS) > 0$,
 \item $F'(1) = -[\uHS(1) - \uLS(1)] = -[(\IHS - \ILS) - \cHS] > 0$,
 \item $F'(q^*) = q^* \cdot (1-q^*) \cdot [\u'_{HS}(q^*) - \u'_{LS}(q^*)] < 0$.
\end{enumerate}
The second inequality follows from the condition \eqref{stationary:N}, and the third inequality is a consequence of the following facts: $q^* \in (0,1)$ and
$$\u'_{HS}(q) - \u'_{LS}(q) = \beta \cdot (\IHS - \ILS) - \cHS < 0 \quad \forall q \in [0,1].$$
Now because $q^*$ is the only asymptotically stable state in $[0,1]$, thus the state $q^*$ is in fact globally asymptotically stable.
\end{proof}


\section{} \label{ApxMNs}
\begin{proof}[{\bf Proof of Corollary \ref{MNs}}]
We denote the right-hand-sides of the system \eqref{evolution:MN} by $f_1(p,q)$ and $f_2(p,q)$ respectively,
i.e.
\begin{align*}
f_1 (p,q) &= p \cdot (1-p) \cdot [\uA(p,q) - \uNA(p,q)]\\
&= p \cdot (1-p) \cdot \bigg[ (1-\beta) \cdot (\IA + q \cdot \IE - \INA) + \beta \cdot p \cdot (\IA + q \cdot \IE - \INA)\\
&- \frac{\beta}{1+m}\big( q \cdot ( \IHS - \ILS) + (\ILS - p \cdot m \cdot A - \IA) \big) - (\cA - A) \bigg],
\end{align*}
\begin{align*}
f_2 (p,q) &= q \cdot (1-q) \cdot [\uHS(p,q) - \uLS(p,q)]\\
&= q \cdot (1-q) \cdot \bigg[(1-\beta) \cdot (\IHS - \ILS) + \frac{\beta \cdot q}{1 + p \cdot m}(\IHS - \ILS) - q \cdot \cHS \bigg].
\end{align*}
We additionally denote
\begin{align*}
h_1 (p,q) &:= \uA(p,q) - \uNA(p,q),\\
h_2 (p,q) &:= \uHS(p,q) - \uLS(p,q).
\end{align*}

Stability of a steady state $(r,s)$ of the equations \eqref{evolution:MN} can be determined by the eigenvalues of the Jacobian matrix of the system (see the Hartman-Grobman theorem \cite{Arrowsmith1992}). Namely:
\begin{itemize}
\item if all eigenvalues of the Jacobian matrix at the state $(r,s)$ have strictly negative real parts then $(r,s)$ is asymptotically stable,
\item if at least one eigenvalue of the Jacobian matrix at the state $(r,s)$ has strictly positive real part then $(r,s)$ is unstable.
\end{itemize}

The Jacobian matrix of the system \eqref{evolution:MN} has the following form:
$$
 \J(p,q) = \left[\begin{array}{cc}
 \frac{\partial f_1}{\partial p}(p,q) & \frac{\partial f_1}{\partial q}(p,q) \\
 \frac{\partial f_2}{\partial p}(p,q) & \frac{\partial f_2}{\partial q}(p,q)
 \end{array}\right],
$$
where
\begin{align*}
\frac{\partial f_1}{\partial p}(p,q) &= (1-p) \cdot h_1 (p,q) - p \cdot h_1 (p,q) + p \cdot (1-p) \cdot \frac{\partial h_1}{\partial p}(p,q),\\
\frac{\partial f_1}{\partial q}(p,q) &= p \cdot (1-p) \cdot \frac{\partial h_1}{\partial q}(p,q),\\
\frac{\partial f_2}{\partial p}(p,q) &= q \cdot (1-q) \cdot \frac{\partial h_2}{\partial p}(p,q),\\
\frac{\partial f_2}{\partial q}(p,q) &= (1-q) \cdot h_2 (p,q) - q \cdot h_2 (p,q) + q \cdot (1-q) \cdot \frac{\partial h_2}{\partial q}(p,q),
\end{align*}
and
\begin{align*}
\frac{\partial h_1}{\partial p}(p,q) &= \beta \cdot (\IA + q \cdot \IE - \INA) + \frac{m}{1+m} \beta \cdot A,\\
\frac{\partial h_1}{\partial q}(p,q) &= (1-\beta) \cdot \IE + \beta \cdot p \cdot \IE - \frac{\beta}{1+m}(\IHS - \ILS),\\
\frac{\partial h_2}{\partial p}(p,q) &= -\left(\frac{1}{1 + p \cdot m}\right)^2 \cdot \beta \cdot m \cdot q \cdot (\IHS - \ILS),\\
\frac{\partial h_2}{\partial q}(p,q) &= \frac{\beta}{1 + p \cdot m}(\IHS - \ILS) - c_{HS}.
\end{align*}

We denote the eigenvalues of the Jacobian matrix by $e_1$ and $e_2$. Let us now compute the eigenvalues in the case of the steady states of the system \eqref{evolution:MN}:

\begin{description}
\item [State $(0,0)$]
$$
 \J(0,0) = \left[\begin{array}{cc}
 h_1 (0,0) & 0 \\
 0 & h_2 (0,0)
 \end{array}\right]
$$

\begin{align*}
e_1 &= h_1 (0,0) = (1-\beta) \cdot (\IA - \INA) - \frac{\beta}{1+m}(\ILS - \IA) - (\cA - A)\\
e_2 &= h_2 (0,0) = (1-\beta) \cdot (\IHS - \ILS)
\end{align*}

Observe that the second eigenvalue is positive, $e_2 > 0$, therefore the state $(0,0)$ is unstable.

\item [State $(0,1)$]
$$
 \J(0,1) = \left[\begin{array}{cc}
 h_1 (0,1) & 0 \\
 0 & -h_2 (0,1)
 \end{array}\right]
$$

\begin{align*}
e_1 &= h_1 (0,1) = (1-\beta) \cdot (\IA + \IE - \INA) - \frac{\beta}{1+m}(\IHS - \IA) - (\cA - A)\\
e_2 &= -h_2 (0,1) = \cHS - (\IHS - \ILS)
\end{align*}

Observe that the second eigenvalue is positive due to inequality \eqref{stationary:N}, $e_2 > 0$, therefore the state $(0,1)$ is unstable.

\item [State $(1,0)$]
$$
 \J(1,0) = \left[\begin{array}{cc}
 -h_1 (1,0) & 0 \\
 0 & h_2 (1,0)
 \end{array}\right]
$$

\begin{align*}
e_1 &= -h_1 (1,0) = \frac{\beta}{1+m}(\ILS - m \cdot A - \IA) + (\cA - A) - (\IA - \INA)\\
e_2 &= h_2 (1,0) = (1-\beta) \cdot (\IHS - \ILS)
\end{align*}

Observe that the second eigenvalue is positive, $e_2 > 0$, therefore the state $(1,0)$ is unstable.

\item [State $(1,1)$]
$$
 \J(1,1) = \left[\begin{array}{cc}
 -h_1 (1,1) & 0 \\
 0 & -h_2 (1,1)
 \end{array}\right]
$$

\begin{align*}
e_1 &= -h_1 (1,1) = \frac{\beta}{1+m}(\IHS - m \cdot A - \IA) + (\cA - A) - (\IA + \IE - \INA)\\
e_2 &= -h_2 (1,1) = \cHS - \left( 1 - \beta + \frac{\beta}{1 + m} \right) \cdot (\IHS - \ILS)
\end{align*}

Because $1 - \beta + \frac{\beta}{1 + m} < 1$ the second eigenvalue is positive due to inequality \eqref{stationary:N}, $e_2 > 0$. Therefore the state $(1,1)$ is unstable.

\item [State $(p^*,0)$]
$$
 \J(p^*,0) = \left[\begin{array}{cc}
 \frac{\partial f_1}{\partial p}(p^*,0) & \frac{\partial f_1}{\partial q}(p^*,0) \\
 0 & h_2(p^*,0)
 \end{array}\right]
$$

\begin{align*}
e_1 &= \frac{\partial f_1}{\partial p}(p^*,0) = (1-p^*) \cdot h_1 (p^*,0) - p^* \cdot h_1 (p^*,0) + p^* \cdot (1-p^*) \cdot \frac{\partial h_1}{\partial p}(p^*,0)\\
&= p^* \cdot (1-p^*) \cdot \left( \beta \cdot (\IA - \INA) + \frac{m}{1+m} \beta \cdot A \right)\\
e_2 &= h_2(p^*,0) = (1-\beta) (\IHS-\ILS)
\end{align*}
(In the computations above it is worth to notice that $h_1 (p^*,0)=0$.)

Observe that both eigenvalues are positive, thus the state $(p^*,0)$ is unstable.

\item [State $(p^{**},1)$]
$$
 \J(p^{**},0) = \left[\begin{array}{cc}
 \frac{\partial f_1}{\partial p}(p^{**},1) & \frac{\partial f_1}{\partial q}(p^{**},1) \\
 0 & -h_2(p^{**},1)
 \end{array}\right]
$$

\begin{align*}
e_1 &= \frac{\partial f_1}{\partial p}(p^{**},1) = (1-p^{**}) \cdot h_1 (p^{**},1) - p^{**} \cdot h_1 (p^{**},1) + p^{**} \cdot (1-p^{**}) \cdot \frac{\partial h_1}{\partial p}(p^{**},1)\\
&= p^{**} \cdot (1-p^{**}) \cdot \left( \beta \cdot (\IA + \IE - \INA) + \frac{m}{1+m} \beta \cdot A \right)\\
e_2 &= -h_2(p^{**},1) = \cHS - \left( 1-\beta + \frac{\beta}{1+ p^{**} \cdot m} \right) (\IHS-\ILS)
\end{align*}
(In the computations above it is worth to notice that $h_1 (p^{**},1)=0$.)

Observe that both eigenvalues are positive (because $1-\beta + \frac{\beta}{1+ p^{**} \cdot m} < 1$ the positive sign of $e_2$ is a consequence of inequality \eqref{stationary:N}), thus the state $(p^{**},1)$ is unstable.
\end{description}
\end{proof}


\section{} \label{ApxMNeq}

\begin{proof}[{\bf Proof of Claim \ref{AS}}]

A steady state of a dynamical system is asymptotically stable if both eigenvalues of its Jacobian matrix are negative.

Jacobian matrix for the state $(0,q^*)$ has the following form
$$
 \J(0,q^*) = \left[\begin{array}{cc}
 h_1 (0,q^*) & 0 \\
 \frac{\partial f_2}{\partial p}(0,q^*) & \frac{\partial f_2}{\partial q}(0,q^*)
 \end{array}\right]
$$
and its eigenvalues are given by:
\begin{align*}
e_1 &= h_1 (0,q^*) = (1-\beta) \cdot (\IA - \INA) - \frac{\beta}{1+m}(\ILS - \IA) - (\cA - A)\\
 &- q^* \cdot \left( \frac{\beta}{1+m}(\IHS - \ILS) - (1-\beta) \cdot \IE \right)\\
e_2 &= \frac{\partial f_2}{\partial q}(0,q^*) = (1-q^*) \cdot h_2 (0,q^*) - q^* \cdot h_2 (0,q^*) + q^* \cdot (1-q^*) \cdot \frac{\partial h_2}{\partial q}(0,q^*)\\
 & = q^* \cdot (1-q^*) \cdot [ \beta \cdot (\IHS - \ILS) - c_{HS} ]
\end{align*}
(In the computations above it is worth to notice that $h_2 (0,q^*)=0$.)

By inequality \eqref{stationary:N} we have that the second eigenvalue is negative, $e_2<0$. Thus we determine when $e_1<0$:
\begin{align*}
e_1 < 0 \Leftrightarrow\ &(1-\beta) \cdot (\IA - \INA) - \frac{\beta}{1+m}(\ILS - \IA) - (\cA - A) \\
 &- q^* \cdot \left( \frac{\beta}{1+m}(\IHS - \ILS) - (1-\beta) \cdot \IE \right) < 0\\
 \Leftrightarrow A &<\: \cA + \frac{\beta}{1+m}(\ILS - \IA) - (1-\beta) \cdot (\IA - \INA) \\
 &+ q^* \cdot \left( \frac{\beta}{1+m}(\IHS - \ILS) - (1-\beta) \cdot \IE \right).
\end{align*}

\vspace{0.3cm}

Jacobian matrix for the state $(1,q^{**})$ has the following form
$$
 \J(1,q^{**}) = \left[\begin{array}{cc}
 -h_1 (1,q^{**}) & 0 \\
 \frac{\partial f_2}{\partial p}(1,q^{**}) & \frac{\partial f_2}{\partial q}(1,q^{**})
 \end{array}\right]
$$
and it has eigenvalues:
\begin{align*}
e_1 &= -h_1 (1,q^{**}) = \frac{\beta}{1+m}(\ILS - m \cdot A - \IA) + (\cA - A) - (\IA - \INA)\\
 &+ q^{**} \cdot \left( \frac{\beta}{1+m}(\IHS - \ILS) - \IE \right)\\
e_2 &= \frac{\partial f_2}{\partial q}(1,q^{**}) = (1-q^{**}) \cdot h_2 (1,q^{**}) - q^{**} \cdot h_2 (1,q^{**}) + q^{**} \cdot (1-q^{**}) \cdot \frac{\partial h_2}{\partial q}(1,q^{**})\\
 & = q^{**} \cdot (1-q^{**}) \cdot \left( \frac{\beta}{1 + m}(\IHS - \ILS) - c_{HS} \right).
\end{align*}
(In the computations above it is worth to notice that $h_2 (1,q^{**})=0$.)

By inequality \eqref{stationary:N} we have that the second eigenvalue is negative, $e_2<0$. Therefore we determine when when the first eigenvalue is negative:
\begin{align*}
e_1 < 0 \Leftrightarrow\ &\frac{\beta}{1+m}(\ILS - m \cdot A - \IA) + (\cA - A) - (\IA - \INA)\\
 &+ q^{**} \cdot \left( \frac{\beta}{1+m}(\IHS - \ILS) - \IE \right) < 0 \\
 \Longleftrightarrow A &>\: \frac{1}{\left( 1+\frac{m}{1+m}\beta \right)} \left[\cA + \frac{\beta}{1+m}(\ILS - \IA) - (\IA - \INA) \right. \\
 &\left. +\: q^{**} \cdot \left( \frac{\beta}{1+m}(\IHS - \ILS) - \IE \right) \right].
\end{align*}

\end{proof}

\begin{proof}[{\bf Proof of Corollary \ref{A**A*}}]

For proving that $A^{**} < A^*$ it is enough to show that $\Big($because $\frac{1}{\left( 1+\frac{m}{1+m}\beta \right)}<1 \Big)$
\begin{align*}
&\cA + \frac{\beta}{1+m}(\ILS - \IA) - (1-\beta) \cdot (\IA - \INA)
 + q^* \cdot \left( \frac{\beta}{1+m}(\IHS - \ILS) - (1-\beta) \cdot \IE \right)\\
 &> \cA + \frac{\beta}{1+m}(\ILS - \IA) - (\IA - \INA)
 + q^{**} \cdot \left(  \frac{\beta}{1+m}(\IHS - \ILS) - \IE \right).
\end{align*}
The last inequality reduces to the following one
\begin{equation} \label{A**A*proof1}
\beta \cdot (\IA - \INA) + \IE \cdot (q^{**} - (1-\beta) \cdot q^*) + \frac{\beta}{1+m}(\IHS - \ILS) \cdot (q^* - q^{**}) > 0.
\end{equation}
Observe that $\beta \cdot (\IA - \INA) > 0$ and by \eqref{q*q**} we have that $\frac{\beta}{1+m}(\IHS - \ILS) \cdot (q^* - q^{**}) > 0$. Thus for proving \eqref{A**A*proof1} it is sufficient to show that $(q^{**} - (1-\beta) \cdot q^*) > 0$. By using \eqref{q*} and \eqref{q**} we have that
\begin{align*}
q^{**} > (1-\beta) \cdot q^* \Longleftrightarrow
\frac{(1-\beta) \cdot (\IHS-\ILS)}{\cHS- \frac{\beta}{1+m} \cdot (\IHS-\ILS)}
> \frac{(1-\beta)^2 \cdot (\IHS-\ILS)}{\cHS- \beta \cdot (\IHS-\ILS)}.
\end{align*}
By \eqref{stationary:N} we know that in the last inequality both denominators are positive. Thus the last inequality is equivalent to
$$
 \cHS - \beta \cdot (\IHS-\ILS) > (1-\beta) \cdot \left( \cHS - \frac{\beta}{1+m} \cdot (\IHS-\ILS) \right),
$$
which reduces to
\begin{equation}\label{A**A*proof2}
 \cHS - \frac{\beta + m}{1 + m} (\IHS - \ILS) > 0.
\end{equation}
Because $\frac{\beta + m}{1 + m} < 1$ we obtain by \eqref{stationary:N} that the inequality \eqref{A**A*proof2} holds true.

\vspace{0.3cm}

For the proof that $A^* < \cA$ observe that this inequality is equivalent to
$$
 \frac{\beta}{1 + m} (\ILS - \IA) - (1-\beta) \cdot (\IA-\INA) + q^* \left( \frac{\beta}{1 + m} (\IHS - \ILS) - (1-\beta) \cdot \IE \right) < 0.
$$
In order to show that the last inequality holds true observe that, by using \eqref{IHS-ILSIq} and because $q^* < 1$, we can estimate its left-hand-side:
\begin{align}
 &\frac{\beta}{1 + m} (\ILS - \IA) - (1-\beta) \cdot (\IA-\INA) + q^* \left( \frac{\beta}{1 + m} (\IHS - \ILS) - (1-\beta) \cdot \IE \right) \nonumber \\
 &< \frac{\beta}{1 + m} (\ILS - \IA) - (1-\beta) \cdot (\IA-\INA) + \frac{\beta}{1 + m} (\IHS - \ILS) - (1-\beta) \cdot \IE \nonumber \\
 &= \frac{\beta}{1 + m} (\IHS - \IA) - (1-\beta) \cdot (\IA + \IE -\INA). \label{A**A*proof3}
\end{align}
Finally notice that the expression in \eqref{A**A*proof3} is negative by \eqref{no_assimilation_cost}.

\end{proof}

\begin{proof}[{\bf Proof of Lemma \ref{CaseI}}]

The solutions of the equations $\uA(p,q)=\uNA(p,q)$ and $\uHS(p,q)=\uLS(p,q)$ with respect to variable $q$ are respectively:
\begin{equation}\label{uA=uNA}
q = \frac{\frac{\beta}{1+m} (\ILS - p \cdot m \cdot A - \IA) + (\cA-A) - (1-\beta + \beta \cdot p) \cdot (\IA - \INA)}{(1-\beta + \beta \cdot p) \cdot \IE - \frac{\beta}{1+m} (\IHS-\ILS)},
\end{equation}
and
\begin{equation}\label{uHS=uLS}
q = \frac{(1-\beta) \cdot (\IHS-\ILS)}{\cHS - \frac{\beta}{1 + p \cdot m}(\IHS-\ILS)}.
\end{equation}
By equating the right-hand-sides of \eqref{uA=uNA} and \eqref{uHS=uLS} and after rearranging we obtain the
following equation:
$$
 a \cdot p^2 + b \cdot p + c = 0,
$$
where
$$
 a = \beta \cdot m \cdot \left[ (1-\beta) \cdot (\IHS-\ILS) \cdot \IE + \cHS \cdot \left( \IA - \INA + \frac{m}{1+m} A \right) \right],
$$ 
\begin{align*}
 b &= \beta \cdot (1-\beta) \cdot (\IHS-\ILS) \cdot \IE \\
 &+ \beta \cdot \big( \cHS - \beta \cdot (\IHS-\ILS) \big) \cdot \left( \IA-\INA+\frac{m}{1+m}A \right) \\
 &- (1-\beta) \cdot m \cdot (\IHS-\ILS) \cdot \left( \frac{\beta}{1+m} (\IHS-\ILS) - (1-\beta) \cdot \IE  \right) \\
 &- m \cdot \cHS \cdot \left( \frac{\beta}{1+m} (\ILS-\IA) + (\cA-A) - (1-\beta) \cdot (\IA-\INA) \right),
\end{align*}
\begin{align*}
c &= \big( \cHS-\beta \cdot (\IHS-\ILS) \big) \cdot \left( (1-\beta) \cdot (\IA-\INA) - \frac{\beta}{1+m} (\ILS-\IA) - \cA + A \right) \\
 &-(1-\beta) \cdot (\IHS-\ILS) \cdot \left( \frac{\beta}{1+m} (\IHS-\ILS) - (1-\beta) \cdot \IE \right).
\end{align*}
We denote by $(p_1, q_1)$, $(p_2,q_2)$ the solutions of the system of equations (I). Without loss of generality we can assume that $p_1 \leqslant p_2$. Then by the Vieta's formulas we have that
\begin{align}
\begin{split}\label{Vieta}
p_1 \cdot p_2 &= \frac{c}{a}, \\
p_1 + p_2 &= - \frac{b}{a}. 
\end{split}
\end{align}
It is evident that $a>0$. Let us determine the sings of $b$ and $c$. For this task we treat $b$ and $c$ as functions of the parameter $A$.

We first determine the sign of $c(\cdot)$. By \eqref{stationary:N} we have that $c'(A) = \cHS-\beta \cdot (\IHS-\ILS) > 0$. The last inequality implies that the function $c(\cdot)$ is strictly increasing. Moreover
\begin{align*}
c(A^*) &= \big( \cHS-\beta \cdot (\IHS-\ILS) \big) \cdot \left( (1-\beta) \cdot (\IA-\INA) - \frac{\beta}{1+m} (\ILS-\IA) - \cA + A^* \right) \\
 &-(1-\beta) \cdot (\IHS-\ILS) \cdot \left( \frac{\beta}{1+m} (\IHS-\ILS) - (1-\beta) \cdot \IE \right)\\
 &= q^* \cdot \big( \cHS-\beta \cdot (\IHS-\ILS) \big) \cdot \left( \frac{\beta}{1+m} (\IHS-\ILS) - (1-\beta) \cdot \IE \right) \\
 &-(1-\beta) \cdot (\IHS-\ILS) \cdot \left( \frac{\beta}{1+m} (\IHS-\ILS) - (1-\beta) \cdot \IE \right)\\
 &= \big( \cHS-\beta \cdot (\IHS-\ILS) \big) \cdot \left( \frac{\beta}{1+m} (\IHS-\ILS) - (1-\beta) \cdot \IE \right) \\
 &\cdot \left[ q^* - \frac{(1-\beta) \cdot (\IHS-\ILS)}{\cHS-\beta \cdot (\IHS-\ILS)} \right] \\
 &= 0
\end{align*}
Thus
\begin{align}
\begin{split}\label{Vietac}
c(A) < 0\quad &\text{for} \quad A < A^* \\
c(A) = 0\quad &\text{for} \quad A = A^* \\
c(A) > 0\quad &\text{for} \quad A > A^*
\end{split}
\end{align}

We now determine the sign of $b(\cdot)$. By \eqref{stationary:N} we have that 
$b'(A) = \frac{\beta \cdot m}{1+m}\big( \cHS - \beta \cdot (\IHS-\ILS) \big) + m \cdot \cHS > 0$. Thus function $b(\cdot)$ is strictly increasing. Moreover
\begin{align}
 b(A^*) &= \beta \cdot (1-\beta) \cdot (\IHS-\ILS) \cdot \IE \nonumber \\
 &+ \beta \cdot \big( \cHS - \beta \cdot (\IHS-\ILS) \big) \cdot \left( \IA-\INA+\frac{m}{1+m}A^* \right) \nonumber \\
 &- (1-\beta) \cdot m \cdot (\IHS-\ILS) \cdot \left( \frac{\beta}{1+m} (\IHS-\ILS) - (1-\beta) \cdot \IE  \right) \nonumber \\
 &+ m \cdot \cHS \cdot q^* \cdot \left( \frac{\beta}{1+m} (\IHS-\ILS) - (1-\beta) \cdot \IE \right) \nonumber \\
 &= \beta \cdot (1-\beta) \cdot (\IHS-\ILS) \cdot \IE \nonumber \\
 &+ \beta \cdot \big( \cHS - \beta \cdot (\IHS-\ILS) \big) \cdot \left( \IA-\INA+\frac{m}{1+m}A^* \right) \nonumber \\
 &+ m \cdot \left( \frac{\beta}{1+m} (\IHS-\ILS) - (1-\beta) \cdot \IE  \right) \cdot \big( \cHS \cdot q^* - (1-\beta) \cdot (\IHS-\ILS) \big) \nonumber \\
 &= \beta \cdot (1-\beta) \cdot (\IHS-\ILS) \cdot \IE \label{Vietab1} \\
 &+ \beta \cdot \big( \cHS - \beta \cdot (\IHS-\ILS) \big) \cdot \left( \IA-\INA+\frac{m}{1+m}A^* \right) \label{Vietab2} \\
 &+ m \cdot \beta \cdot (1-\beta) \cdot \left( \frac{\beta}{1+m} (\IHS-\ILS) - (1-\beta) \cdot \IE  \right) \frac{(\IHS-\ILS)^2}{\cHS - \beta \cdot (\IHS-\ILS)}. \label{Vietab3}
\end{align}
Now the expression \eqref{Vietab1} is positive, by \eqref{stationary:N} the expression \eqref{Vietab2} is positive and by the inequality \eqref{IHS-ILSIq} the expression \eqref{Vietab3} is positive. Therefore
\begin{equation}\label{Vietab}
b(A) > 0 \quad \text{for} \quad A \geqslant A^*.
\end{equation}

By using \eqref{Vieta}, \eqref{Vietac} and \eqref{Vietab} we derive the following conclusions:
\begin{description}
\item[Case $A<A^*$.] Because $p_1 \cdot p_2 < 0$ we have that $p_1 < 0$ and $p_2 > 0$. Because $p_1 < 0$ we obtain that $(p_1,q_1) \notin [0,1] \times [0,1]$.

\item[Case $A=A^*$.] Because $p_1 \cdot p_2 = 0$ and $p_1 + p_2 < 0$ we have that $p_1 < 0$ and $p_2 = 0$. Because $p_1 < 0$ we obtain that $(p_1,q_1) \notin [0,1] \times [0,1]$.

\item[Case $A>A^*$.] Because $p_1 \cdot p_2 > 0$ and $p_1 + p_2 < 0$ we have that $p_1 < 0$ and $p_2 < 0$. Therefore $(p_1,q_1) \notin [0,1] \times [0,1]$ and $(p_2,q_2) \notin [0,1] \times [0,1]$.
\end{description}
\end{proof}

\begin{proof}[{\bf Proof of Corollary \ref{olp,olq}}]
The proof of thesis 1. is straightforward. Indeed, we have that $q^{**} \leqslant \ol{q} \leqslant q^*$ if and only if
\begin{align*}
&\frac{(1-\beta) \cdot (\IHS-\ILS)}{\cHS - \frac{\beta}{1 + m} (\ILS-\ILS)} \leqslant \frac{(1-\beta) \cdot (\IHS-\ILS)}{\cHS - \frac{\beta}{1 + \ol{p} \cdot m} (\ILS-\ILS)} \leqslant \frac{(1-\beta) \cdot (\IHS-\ILS)}{\cHS - \beta \cdot (\ILS-\ILS)} \\
\Longleftrightarrow\ &\cHS - \frac{\beta}{1 + m} (\ILS-\ILS) \geqslant \cHS - \frac{\beta}{1 + \ol{p} \cdot m} (\ILS-\ILS) \geqslant \cHS - \beta \cdot (\ILS-\ILS) \\
\Longleftrightarrow\ &\frac{\beta}{1 + m} (\ILS-\ILS) \leqslant \frac{\beta}{1 + \ol{p} \cdot m} (\ILS-\ILS) \leqslant \beta \cdot (\ILS-\ILS) \\
\Longleftrightarrow\ &1+m \geqslant 1 + \ol{p} \cdot m \geqslant 1 
\Longleftrightarrow 1 \geqslant \ol{p} \geqslant 0.
\end{align*}

For the proof of thesis 2. we compute the Jacobian matrix in the state $(\ol{p},\ol{q})$ (we use the notation from Appendix \ref{ApxMNs})
$$
 \J(\ol{p},\ol{q}) = \left[\begin{array}{cc}
 \frac{\partial f_1}{\partial p}(\ol{p},\ol{q}) & \frac{\partial f_1}{\partial q}(\ol{p},\ol{q}) \\
 \frac{\partial f_2}{\partial p}(\ol{p},\ol{q}) & \frac{\partial f_2}{\partial q}(\ol{p},\ol{q})
 \end{array}\right],
$$
where (notice that $h_1 (\ol{p},\ol{q}) = h_2 (\ol{p},\ol{q}) = 0$):
\begin{align*}
\frac{\partial f_1}{\partial p}(\ol{p},\ol{q}) &= \ol{p} \cdot (1 - \ol{p}) \frac{\partial h_1}{\partial p}(\ol{p},\ol{q}) = \ol{p} \cdot (1 - \ol{p}) \left( \beta \cdot (\IA - \INA) + \beta \cdot  \ol{q} \cdot \IE + \frac{m}{1+m} \beta \cdot A \right),\\
\frac{\partial f_1}{\partial q}(\ol{p},\ol{q}) &= \ol{p} \cdot (1 - \ol{p}) \frac{\partial h_1}{\partial q}(\ol{p},\ol{q}) = \ol{p} \cdot (1 - \ol{p}) \left( (1-\beta) \cdot \IE + \beta \cdot \ol{p} \cdot \IE - \frac{\beta}{1+m}(\IHS - \ILS) \right),\\
\frac{\partial f_2}{\partial p}(\ol{p},\ol{q}) &= \ol{q} \cdot (1 - \ol{q}) \frac{\partial h_2}{\partial p}(\ol{p},\ol{q}) = \ol{q} \cdot (1 - \ol{q}) \left(\frac{-1}{(1 + \ol{p} \cdot m)^2} \cdot \beta \cdot m \cdot \ol{q} \cdot (\IHS - \ILS) \right),\\
\frac{\partial f_2}{\partial q}(\ol{p},\ol{q}) &= \ol{q} \cdot (1 - \ol{q}) \frac{\partial h_2}{\partial q}(\ol{p},\ol{q}) = \ol{q} \cdot (1 - \ol{q}) \left( \frac{\beta}{1 + \ol{p} \cdot m}(\IHS - \ILS) - c_{HS} \right).
\end{align*}
Now
$$
 \det( \J(\ol{p},\ol{q})) = \ol{p} \cdot (1 - \ol{p}) \cdot \ol{q} \cdot (1 - \ol{q}) \cdot \left[ \frac{\partial h_1}{\partial p}(\ol{p},\ol{q}) \cdot \frac{\partial h_2}{\partial q}(\ol{p},\ol{q}) - \frac{\partial h_1}{\partial q}(\ol{p},\ol{q}) \cdot \frac{\partial h_2}{\partial p}(\ol{p},\ol{q}) \right].
$$
Let us rewrite the expression in the square brackets above
\begin{align}
&\frac{\partial h_1}{\partial p}(\ol{p},\ol{q}) \cdot \frac{\partial h_2}{\partial q}(\ol{p},\ol{q}) - \frac{\partial h_1}{\partial q}(\ol{p},\ol{q}) \cdot \frac{\partial h_2}{\partial p}(\ol{p},\ol{q}) \nonumber \\
&= \left( \beta \cdot (\IA - \INA) + \frac{m}{1+m} \beta \cdot A \right) \cdot \left( \frac{\beta}{1 + \ol{p} \cdot m}(\IHS - \ILS) - c_{HS} \right) \nonumber \\
&+ \left( (1-\beta) \cdot \IE - \frac{\beta}{1+m}(\IHS - \ILS) \right) \frac{\beta \cdot m \cdot \ol{q}}{(1 + \ol{p} \cdot m)^2} (\IHS - \ILS) \nonumber \\
&+ \beta \cdot  \ol{q} \cdot \IE \cdot \left( \frac{\beta}{1 + \ol{p} \cdot m}(\IHS - \ILS) - c_{HS} \right) + \frac{\beta^2 \cdot m \cdot \ol{p} \cdot \ol{q}}{(1 + \ol{p} \cdot m)^2} (\IHS - \ILS) \cdot \IE \nonumber \\
&= \left( \beta \cdot (\IA - \INA) + \frac{m}{1+m} \beta \cdot A \right) \cdot \left( \frac{\beta}{1 + \ol{p} \cdot m}(\IHS - \ILS) - c_{HS} \right) \label{olpolq1} \\
&+ \left( (1-\beta) \cdot \IE - \frac{\beta}{1+m}(\IHS - \ILS) \right) \frac{\beta \cdot m \cdot \ol{q}}{(1 + \ol{p} \cdot m)^2} (\IHS - \ILS) \label{olpolq2} \\
&+ \beta \cdot  \ol{q} \cdot \IE \cdot \left( \beta \frac{1+\frac{\ol{p} \cdot m}{1 + \ol{p} \cdot m}}{1 + \ol{p} \cdot m} (\IHS-\ILS) - \cHS \right) \label{olpolq3}
\end{align}
By \eqref{stationary:N} we have that $\left( \frac{\beta}{1 + \ol{p} \cdot m}(\IHS - \ILS) - c_{HS} \right) < 0$ and $\left( \beta \frac{1+\frac{\ol{p} \cdot m}{1 + \ol{p} \cdot m}}{1 + \ol{p} \cdot m} (\IHS-\ILS) - \cHS \right) < 0$, thus the expressions \eqref{olpolq1} and \eqref{olpolq3} are negative. Moreover by \eqref{IHS-ILSIq} the expression in \eqref{olpolq2} is negative. Therefore
\begin{equation} \label{detolpolq}
\det\big( \J(\ol{p},\ol{q}) \big) < 0.
\end{equation}
Let us denote by $e_1, e_2$ the eigenvalues of $J(\ol{p},\ol{q})$. By \eqref{detolpolq} we have that
$e_1 \cdot e_2 = \det( \J(\ol{p},\ol{q}) < 0$. Consequently one eigenvalue is positive and the other eigenvalue is negative. As a result the state $(\ol{p},\ol{q})$ is unstable.
\end{proof}


\section{} \label{ApxSW}

\begin{proof}[{\bf Proof of Claim \ref{full_assimilationN}}]

Because $\uHS(0,q^*) = \uLS(0,q^*)$ and $\uHS(1,q^{**}) = \uLS(1,q^{**})$ we have by \eqref{SWN} that
\begin{align*}
 \SWN(0,q^*,0) &= N \cdot \uLS(0,q^*)_{\big| A=0} \\
 &= N \cdot \left[ (1-\beta) \cdot ( \ILS + q^* \cdot \IE) - \beta \cdot q^* \cdot (\IHS - \ILS) \right] \\
 &= N \cdot \left[ q^* \cdot \big( (1-\beta) \cdot \IE - \beta \cdot (\IHS - \ILS) \big) + (1-\beta) \cdot \ILS \right],
\end{align*}
and
\begin{align*}
 \SWN(1,q^{**},A^*) &= N \cdot \uLS(1,q^{**})_{\big| A=A^*} \\
 &= N \cdot \left[ (1-\beta) \cdot ( \ILS + q^{**} \cdot \IE - m \cdot A^*) - \frac{\beta \cdot q^{**}}{1+m} (\IHS - \ILS) \right] \\
 &= N \cdot \Bigg[ q^{**} \cdot \left( (1-\beta) \cdot \IE - \frac{\beta}{1+m} (\IHS - \ILS) \right) + (1-\beta) \cdot \ILS \\
 &\qquad\quad - (1-\beta) \cdot m \cdot A^* \Bigg].
\end{align*}
Thus $\SWN(1,q^{**},A^*) > \SWN(0,q^*,0)$ if and only if
\begin{align*}
 &q^{**} \cdot \left( (1-\beta) \cdot \IE - \frac{\beta}{1+m} (\IHS - \ILS) \right) - (1-\beta) \cdot m \cdot A^* \\
 &> q^* \cdot \big( (1-\beta) \cdot \IE - \beta \cdot (\IHS - \ILS) \big),
\end{align*}
or equivalently
\begin{align*}
 (1-\beta) \cdot m \cdot A^* <\: &q^{**} \cdot \left( (1-\beta) \cdot \IE - \frac{\beta}{1+m} (\IHS - \ILS) \right) \\
 &- q^* \cdot \big( (1-\beta) \cdot \IE - \beta \cdot (\IHS - \ILS) \big).
\end{align*}
By using \eqref{q*} and \eqref{q**} we simplify the last inequality to
\begin{equation} \label{full_assimilationN_1}
 (1-\beta) \cdot m \cdot A^* < q^* \cdot q^{**} \cdot \frac{\beta}{1-\beta} \cdot \frac{m}{1+m} \left( \cHS - (1-\beta) \cdot \IE \right).
\end{equation}
Because $A^* = \cA - \ol\cA$ (compare \eqref{A*} with \eqref{olcA}) we obtain form \eqref{full_assimilationN_1} that
$$
 \SWN(1,q^{**},A^*) > \SWN(0,q^*,0) \Longleftrightarrow \cA < \ol{\cA} + q^* \cdot q^{**} \cdot \frac{\beta \cdot \left( \cHS - (1-\beta) \cdot \IE \right)}{(1+m)(1-\beta)^2}.
$$

\end{proof}

\begin{proof}[{\bf Proof of Corollary \ref{full_assimilationM}}]

We have that
\begin{align*}
 \SWM(0,q^*,0) = M \cdot \uNA(0,q^*)_{\big| A=0} = M \cdot (1-\beta) \cdot \INA
\end{align*}
and that
\begin{align*}
 \SWM(1,q^{**},A^*) &= M \cdot \uA(1,q^{**})_{\big| A=A^*} \\
 &= M \cdot \left[ \left( 1+\frac{m}{1+m}\beta \right) \cdot A^* + (1-\beta) \cdot \IA - \frac{\beta}{1+m} \left( \ILS - \IA \right) - \cA \right. \\
 &\qquad\quad\left. - q^{**} \cdot \left( \frac{\beta}{1+m} (\IHS - \ILS) - (1-\beta) \cdot \IE \right) \right].
\end{align*}
Thus $\SWM(1,q^{**},A^*) > \SWM(0,q^*,0)$ if and only if
\begin{align}
\begin{split}\label{full_assimilationM_1}
 &\left( 1+\frac{m}{1+m}\beta \right) \cdot A^* + (1-\beta) \cdot (\IA-\INA) - \frac{\beta}{1+m} \left( \ILS - \IA \right) - \cA \\
 &- q^{**} \cdot \left( \frac{\beta}{1+m} (\IHS - \ILS) - (1-\beta) \cdot \IE \right) > 0.
\end{split}
\end{align}
By using \eqref{A*} we have that
$$
 (1-\beta) \cdot (\IA-\INA) - \frac{\beta}{1+m} \left( \ILS - \IA \right) - \cA = q^{*} \cdot \left( \frac{\beta}{1+m} (\IHS - \ILS) - (1-\beta) \cdot \IE \right) - A^*
$$
By including this fact in \eqref{full_assimilationM_1} we obtain the following inequality
$$
 \frac{m}{1+m}\beta \cdot A^* >  \left( q^{**} - q^* \right) \cdot \left( \frac{\beta}{1+m} (\IHS - \ILS) - (1-\beta) \cdot \IE \right).
$$
Now the last inequality is satisfied because its left-hand-side is positive (in Section \ref{SW} we assume that $A^*>0$) and the right-hand-side is negative by \eqref{IHS-ILSIq} and \eqref{q*q**}.

\end{proof}

\end{appendices}






\begin{thebibliography}{00}

\bibitem{Akerlof1997} Akerlof, George A. (1997). ``Social Distance and Social Decisions,'' {\it Econometrica} 65(5): 1005-1027.

\bibitem{Arrowsmith1992} Arrowsmith, D. K., Place, C. M. (1992). {\it Dynamical Systems: Differential Equations, Maps, and Chaotic Behaviour.} London: Chapman \& Hall.

\bibitem{BarreiraDaSilvaRocha2013} Barreira Da Silva Rocha, Andr\'e (2013). ``Evolutionary Dynamics of Nationalism and Migration,'' {\it Physica A: Statistical Mechanics and its Applications} 392(15): 3183-3197.

\bibitem{BeizinMoizeau2017} Bezin, Emeline and Moizeau, Fabien (2017). ``Cultural Dynamics, Social Mobility and Urban Segregation'' {\it Journal of Urban Economics} 99: 173-187. 

\bibitem{Borjas2017} Borjas, George J. (2017). ``The Wage Impact of the Marielitos: A Reappraisal,'' {\it ILR Review} 70(5): 1077-1110. 

\bibitem{Borjasetal1992} Borjas, George J., Bronars, Stephen G., and Trejo, Stephen J. (1992). ``Assimilation and the Earnings of Young Internal Migrants,''. {\it Review of Economics and Statistics} 74(1): 170-175.

\bibitem{BossertDAmbrosio2006} Bossert, W. and C. D'Ambrosio (2006). ``Reference Groups and Individual Deprivation,'' {\it Economics Letters}, 90, 421-426.

\bibitem{ChiswickMiller2005} Chiswick, Barry R. and Miller, Paul W. (2005). ``Do Enclaves Matter in Immigrant Adjustment?,'' {\it City \& Community} 4(1): 5-36.

\bibitem{Clark2008} Clark, Andrew E., Frijters, Paul, and Shields, Michael A. (2008). ``Relative Income, Happiness, and Utility: An Explanation for the Easterlin Paradox and Other Puzzles,'' {\it Journal of Economic Literature} 46(1): 95-144.

\bibitem{Cutler2008} Cutler, David M., Glaeser, Edward L., and Vigdor, Jacob L. (2008). ``When are Ghettos Bad? Lessons from Immigrant Segregation in the United States,'' {\it Journal of Urban Economics} 63(3): 759-774.

\bibitem{EbertMoyes2000} Ebert, Udo and Moyes, Patrick (2000). ``An Axiomatic Characterization of Yitzhaki's Index of Individual Deprivation,'' {\it Economics Letters}, 68, 263-270.


\bibitem{Eurostat2018} Eurostat (2018). ``Migration Integration Statistics - at Risk of Poverty and Social Exclusion,'' online publication available at \verb+https://ec.europa.eu/eurostat/+ \verb+statistics-explained/index.php?title=Migrant_integration_statistics_+ \verb+-_at_risk_of_poverty_and_social_exclusion+, accessed 06.06.2019.

\bibitem{FanStark2007} Fan, C. Simon and Stark, Oded (2007). ``A Social Proximity Explanation of the Reluctance to Assimilate,'' {\it Kyklos} 60(1): 55-63.

\bibitem{FogedPeri2016} Foged, Mette and Peri, Giovanni (2016). ``Immigrants' Effect on Native Workers: New Analysis on Longitudinal Data,'' {\it American Economic Journal: Applied Economics} 8(2): 1-34.


\bibitem{Lazear1999} Lazear, Edward P. (1999). ``Culture and Language,'' {\it Journal of Political Economy} 107(S6): S95-S126.

\bibitem{Luttmer2005} Luttmer, Erzo F. P. (2005). ``Neighbors as Negatives: Relative Earnings and Well-Being,'' {\it Quarterly Journal of Economics} 120(3): 963-1002.

\bibitem{McManus1983} McManus, Walter, Gould, William, and Welch, Finis (1983). ``Earnings of Hispanic Men: The Role of English Language Proficiency,'' {\it Journal of Labor Economics} 1(2): 101-130.

\bibitem{Moretti2004} Moretti, Enrico (2004). ``Estimating the Social Return to Higher Education: Evidence from Longitudinal and Repeated Cross-sectional Data,'' {\it Journal of Econometrics} 121(1-2): 175-212.

\bibitem{Noailly2008} Noailly, Jo\"elle (2008). ``Coevolution of Economic and Ecological Systems: An Application to Agricultural Pesticide Resistance,'' {\it Journal of Evolutionary Economics} 18: 1-29.

\bibitem{Rauch1993} Rauch, James E. (1993). ``Productivity Gains from Geographic Concentration of Human Capital: Evidence from the Cities,'' {\it Journal of Urban Economics} 34(3): 380-400.

\bibitem{ShieldsPrice2002} Shields, Michael A. and Price Stephen W. (2002). ``The English Language Fluency and Occupational Success of Ethnic Minority Immigrant Men Living in English Metropolitan Areas,'' {\it Journal of Population Economics} 15(1): 137-160.

\bibitem{StarkBielawskiJakubek2014} Stark, Oded, Bielawski, Jakub and Jakubek, Marcin (2014). ``The Impact of the Assimilation of Migrants on the Well-being of Native Inhabitants: A Theory,'' {\it Journal of Economic Behavior \& Organization} 111(C): 71-78.

\bibitem{StarkJakubek2013} Stark, Oded and Jakubek, Marcin (2013). ``Integration as a Catalyst for Assimilation,'' {\it International Review of Economics and Finance} 28: 62-70.

\bibitem{StarkJakubekSzczygielski2018} Stark, Oded, Jakubek, Marcin and Szczygielski, Krzysztof (2018). ``Community Cohesion and Assimilation Equilibria,'' {\it Journal of Urban Economics} 107: 79-88.

\bibitem{Tainer1988} Tainer, Evelina (1988). ``English language proficiency and the determination of earnings among foreign-born men,'' The Journal of Human Resources 23(1): 108-122.

\bibitem{WalkerSmith2002} Walker, Iain and Smith, Heather J. (2002). {\it Relative Deprivation: Specification, Development, and Integration}. Cambridge: Cambridge University Press.

\bibitem{Yitzhaki1979} Yitzhaki, Shlomo (1979). ``Relative Deprivation and the Gini Coefficient,'' {\it Quarterly Journal of Economics 93(2): 321-324}.


\end{thebibliography}
\end{document}